\documentclass[prd,nofootinbib,onecolumn,preprint,11pt]{revtex4}

\usepackage{graphicx}
\usepackage{bm}
\usepackage{float}
\usepackage{subfigure}
\usepackage{amsmath}
\usepackage{amsfonts}
\usepackage{color}
\usepackage{slashed}
\usepackage{fancyhdr}

\newcommand{\mpl}{M_{\rm pl}}
\newcommand{\GN}{G_{\rm N}}

\newcommand{\dx}[1]{\text{d}#1}
\newcommand{\dd}{\text{d}}

\newcommand{\mma}{\textsf{Mathematica}}

\begin{document}

\title{Scalar Wave Tails in Even Dimensional Weakly Curved Static Newtonian Spacetimes}

\author{Yi-Zen Chu}
\affiliation{
Center for Particle Cosmology, Department of Physics and Astronomy, University of Pennsylvania, Philadelphia, Pennsylvania 19104, USA
}

\begin{abstract}
\noindent In a 4-dimensional (4D) weak field geometry governed by the linearized Einstein's equations and sourced primarily by a static, spatially localized, but otherwise arbitrary mass density $T_{00}$, it is known that the leading order tail part of the Green's functions of the minimally coupled massless scalar, Lorenz gauge photon, and de Donder gauge graviton at late times are not only time translation symmetric, but also space-translation and spherically symmetric. Only the monopole moment of the matter source is responsible for the late-time tail of the Green's functions. We provide evidence, in this paper, that both of these statements will cease to hold for all even dimensions higher than 4. As a consequence, we anticipate that the late time behavior of massless fields propagating in higher even dimensional asymptotically flat spacetimes will exhibit richer phenomenology than their 4D counterparts.
\end{abstract}

\maketitle

\section{Introduction and Motivation}

The tail effect is the phenomenon of fields propagating inside the light cone of their sources. In curved spacetimes, this occurs even for massless fields -- scalars, photons and gravitons, for example. Understanding the tail effect in black hole spacetimes, in particular, is important for modeling the self-force exerted by compact bodies as they orbit and subsequently plunge into super-massive black holes lying at the center of most galaxies. The detailed calculation of such orbital dynamics, in turn, is crucial for predicting gravitational waves (GWs) generated by such Extreme-Mass-Ratio-Inspiral systems, which are expected to be heard by the future space-based GW observatory eLISA/NGO.

A related aspect of the tail effect concerns how a initial field configuration evolves forward in time and disperses to infinity. Such late time behavior of fields propagating in an asymptotically flat geometric background has been a subject of study since the work of Price \cite{Price:1972pw}. Price showed that in a 4D Schwarzschild geometry, if one begins at time $t=0$ with a scalar field configuration that is proportional to the spherical harmonic $Y_\ell^m[\theta,\phi]$, then at late times $t \gg r$ (where $r$ is the radial coordinate), the field would decay as $1/t^{2\ell+3}$.\footnote{Price's work was for a test field propagating in a fixed background Schwarzschild geometry. For recent work on the fully nonlinear (but spherically symmetric) Einstein-(Massless-)Scalar setup, see for e.g., Luk and Oh \cite{Luk:2014sha}.}

It has been argued by Poisson \cite{Poisson:2002jz}, based on work done by Ching et al. \cite{Ching:1994bd,Ching:1995tj}, that this late time tail behavior can be captured by a weak field calculation because -- at least for asymptotically flat geometries -- it is the result of the scattering of the fields off the geometry at large spatial distances from the source of the geometry. Poisson \cite{Poisson:2002jz} then went on to compute the tail part of the minimally coupled massless scalar retarded Green's function in a 4D geometry that solves the linearized Einstein's equations sourced by a static spatially localized but otherwise arbitrary mass density and momentum current. He showed that, for elapsed times larger than the time it takes for a null ray at $r'$ to scatter off the central mass before proceeding to $r$, i.e., for $t-t' > r+r'$ (where $(t,r)$ and $(t',r')$ are the time and radial coordinates of the observer and source respectively), the tail portion of the Green's function becomes space-translation and spherically symmetric (see his eq. (1.4) and eq. \eqref{4D_Tail} below). Moreover, it is sourced entirely by the monopole moment of the mass density. (The scalar Green's function in a static geometry is necessarily time-translation symmetric.) The anticipated $1/t^{2\ell+3}$ behavior is then gotten by employing the Green's function to time-evolve an initial scalar profile and its velocity that are both proportional to $Y_\ell^m[\theta,\phi]$, using the Kirchhoff integral representations. Although Poisson performed a massless scalar calculation, we will in fact explain in appendix \eqref{Section_LateTimeTails_4D} why the late time tails of the photon and graviton Green's functions should yield essentially the same behavior as their scalar cousin.

To gain insight into physical phenomenon it is sometimes useful to understand it in spacetime dimensions different from the 4 we reside in, to understand what aspects of the underlying physical mechanisms are dimension dependent. Such is the case for massless waves in 2 and odd dimensions: even though the momentum vector of a given plane wave mode is null in any dimension, unlike in $(d \geq 4)$ even dimensions, physical sources produce massless particles that propagate inside their light cone. Furthermore, the radiation reaction and self-force problems have recently been studied in arbitrary dimensional Minkowski spacetimes \cite{Birnholtz:2013ffa} and in a 5D black hole spacetime \cite{Beach:2014aba}. It appears that, in odd dimensions, the analog of the Abraham-Lorentz-Dirac self-force in 4D becomes particularly sensitive to the detailed structure of a given compact body under study.

In this paper, we wish to initiate the study of the tail effect in even dimensional $(d \geq 4)$ weakly curved spacetimes. In particular, we wish to argue that the highly symmetric nature of the late time tail portion of the scalar Green's function found by Poisson \cite{Poisson:2002jz}, as well as the monopole moment being its sole contributor, is special to 4 dimensions.\footnote{Note that in a 2D curved geometry, the minimally coupled massless scalar Green's function is the same as that in Minkowski spacetime, due to conformal invariance.} We will consider a weakly curved geometry that solves the linearized Einstein's equations sourced by a static, spatially localized, but otherwise arbitrary mass density $T_{00}[\vec{x}]$ -- i.e., what we shall call a static Newtonian spacetime. (We will not consider a non-trivial momentum density $T_{0i}$.) We shall then witness that the late time tail part of the minimally coupled massless retarded Green's function in such a geometry, for even $d > 4$, does not have any obvious symmetries. This is because, unlike the $d=4$ case, the monopole and all higher multipole moments contribute to it.

In section \eqref{Section_Setup} we will invoke the perturbation theory developed in \cite{Chu:2011ip} to write down the general formula for the minimally coupled massless scalar Green's function up to first order in metric perturbations. In section \eqref{Section_Newtonian}, we will specialize to a static Newtonian spacetime and solve for the tail part of the scalar Green's function. We conclude in section \eqref{Section_Conclusions}. In appendix \eqref{Section_ProlateEllipsoidalHarmonicExpansion} we derive the prolate ellipsoidal harmonic expansion of $|\vec{x}-\vec{x}'|^{2-D}$ for $(D \geq 3)$-spatial dimensions; this result plays a key role in the analysis of section \eqref{Section_Newtonian}. In appendix \eqref{Section_LateTimeTails_4D}, we explain why the late time tails of photons and gravitons in a 4D weakly curved asymptotically flat static geometry are only sensitive to the mass monopole of the source responsible for the weak field geometry itself.

\section{Setup and Generalities}
\label{Section_Setup}

We will work in $d=4+2n$ spacetime dimensions, where $n=0,1,2,3,\dots$, and assume the geometry is weakly curved; in pseudo-Cartesian coordinates $x^\mu \equiv (t,\vec{x})$, we have
\begin{align}
\label{WeakFieldGeometry}
g_{\mu\nu} = \eta_{\mu\nu} + h_{\mu\nu}, \qquad |h_{\mu\nu}| \ll 1 .
\end{align}
Let us denote the flat spacetime minimally coupled massless scalar retarded Green's function by $\overline{G}_{x,x'}$, where $x'$ is interpreted as the location of its spacetime point source and $x$ as the location of some observer. In \cite{Chu:2011ip} it has been shown that the $\mathcal{O}[h]$ accurate Green's function $G_{x,x'}$, obeying 
\begin{align}
\Box_x G_{x,x'} = \Box_{x'} G_{x,x'} = \frac{\delta^{(d)}[x-x']}{\sqrt[4]{|g[x]g[x']|}} ,
\end{align}
reads
\begin{align}
G_{x,x'} = \overline{G}_{x,x'} + (G|1)_{x,x'} + \mathcal{O}[h^2],
\end{align}
where
\begin{align}
\label{deltaG_Start}
(G|1)_{x,x'}
= \partial_\alpha \partial_{\beta'} 
	\int \dd^d x'' \overline{G}_{x,x''} \left( \frac{1}{2} h'' \eta^{\alpha\beta} - h^{\alpha''\beta''} \right) \overline{G}_{x'',x'} ,
	\qquad h'' \equiv \eta^{\mu\nu} h_{\mu\nu}[x''] .
\end{align}
This formula actually holds for any $d > 2$. Now, the $d=4+2n$ dimensional flat Minkowski scalar Green's function can be generated from its 4D counterpart via
\begin{align}
\label{G_EvenD_Flat}
\overline{G}_{4+2n}[t-t',R \equiv |\vec{x}-\vec{x}'|]
= \left(-\frac{1}{2\pi R} \partial_R \right)^n \frac{\delta[t-t'-R]}{4\pi R}
\equiv \frac{1}{(2\pi)^n} \sum_{\ell=0}^n A_\ell^{(n)} \frac{\delta_{(\ell)}[t-t'-R]}{4\pi R^{2n+1-\ell}} ,
\end{align}
where $\delta_{(\ell)}[\xi] \equiv \partial_\xi^\ell \delta[\xi]$ and the second equality is meant as a definition of the constants $A_\ell^{(n)}$. The central result of this section is that the first order in $h$ piece of the Green's function $G_{x,x'}$ is
{\allowdisplaybreaks
\begin{align}
\label{OrderhGreensFunction_MasslessScalar}
\left(G|1\right)_{x,x'} &=
\frac{1}{2(2\pi)^{2n}} \sum_{\ell_1,\ell_2=0}^{n}  A_{\ell_1}^{(n)} A_{\ell_2}^{(n)} 
	\partial_\alpha \partial_{\beta'} \partial_t^{\ell_1} \partial_{-t'}^{\ell_2} \\
&\times
	\Bigg\{ 
	\Theta[t-t'] \Theta[\bar{\sigma}] \left( \frac{\bar{\sigma}}{2} \right)^n 
	\int_{-1}^{+1} \dd c'' \left(1-c''^2\right)^n \int_{\mathbb{S}^{2n+1}} \dd\Omega_{2n+1}^{(\widehat{n})}
		\frac{\left(\frac{t-t'}{2}-\frac{R}{2}c''\right)^{\ell_1} \left(\frac{t-t'}{2}+\frac{R}{2}c''\right)^{\ell_2}}{(4\pi)^2 \left(\left(\frac{t-t'}{2}\right)^2 - \left(\frac{R}{2}\right)^2 c''^2\right)^{2n}} \nonumber\\
&\qquad\qquad\qquad\times 
H^{\alpha''\beta''}\left[ \frac{1}{2}\left(t+t'+Rc''\right),
\frac{\vec{x}+\vec{x}'}{2} + \sqrt{\frac{\bar{\sigma}}{2}} \sqrt{1-c''^2} \widehat{n}\left[\widehat{\Omega}_{2n+1}\right] + \frac{t-t'}{2} c'' \widehat{z} \right]
\Bigg\}  \nonumber
\end{align}}
where
\begin{align}
H^{\alpha''\beta''} \equiv \frac{1}{2} h'' \eta^{\alpha\beta} - h^{\alpha''\beta''}, \qquad
		\bar{\sigma} \equiv \frac{(t-t')^2-R^2}{2}, \qquad R \equiv |\vec{x}-\vec{x}'| .
\end{align}
The first argument of $H^{\alpha''\beta''}$ in eq. \eqref{OrderhGreensFunction_MasslessScalar} is time and the second argument is the spatial coordinates; $\widehat{z} \equiv (\vec{x}-\vec{x}')/R$ is the unit vector pointing from the source at $\vec{x}'$ to the observer at $\vec{x}$; while $\widehat{n}$ is the unit radial spatial vector, parametrized by the $2n+1$ angles represented by the collective variable $\widehat{\Omega}_{2n+1}$, that is orthogonal to $\widehat{z}$ (i.e., $\widehat{n} \cdot \widehat{z} = 0$); the $\int_{\mathbb{S}^{2n+1}} \dd\Omega_{2n+1}^{(\widehat{n})}$ is the spherically symmetric solid angle integral over $\mathbb{S}^{2n+1}$, with respect to $\widehat{\Omega}_{2n+1}$.

The tail part of the Green's function begins at order $h$, and its general formula is the term in eq. \eqref{OrderhGreensFunction_MasslessScalar} with no derivatives acting on the $\Theta[t-t'] \Theta[\bar{\sigma}]$. Its interpretation is similar in spirit to that of its $d=4$ cousin: the leading order tail effect is due to the collection of null scalar rays emanating from $\vec{x}'$, scattering off the geometry at all possible locations given by $\vec{x}'' = (\vec{x}+\vec{x}')/2 + \vec{x}'''$ (see equations \eqref{ProlateEliipsoidalIntegrationCoordinates_IofII} and \eqref{ProlateEliipsoidalIntegrationCoordinates_IIofII} below), before hitting the observer at $\vec{x}$. Because the paths between the scattering are null, the sum of their spatial distances must be equal to the elapsed time, i.e., giving us the prolate ellipsoid $t-t' = |\vec{x}-\vec{x}''| + |\vec{x}'-\vec{x}''|$, whose focii are at the observer $\vec{x}$ and source $\vec{x}'$ locations. In odd spacetime dimensions, this scattering picture will be complicated by the presence of a non-zero tail already in flat spacetime, so the additional tail contribution from a metric perturbation would not solely be the result of null rays scattering off the geometry, but also by timelike rays doing the same.

{\bf Derivation of eq. \eqref{OrderhGreensFunction_MasslessScalar}} \qquad By expressing the derivatives on the $\delta$-functions in eq. \eqref{G_EvenD_Flat} as derivatives with respect to the observer or emission time, and inserting eq. \eqref{G_EvenD_Flat} into eq. \eqref{deltaG_Start},
\begin{align}
\left(G|1\right)_{x,x'} &=
\frac{1}{(2\pi)^{2n}} \sum_{\ell_1,\ell_2=0}^{n}  A_{\ell_1}^{(n)} A_{\ell_2}^{(n)} 
	\partial_\alpha \partial_{\beta'} \partial_t^{\ell_1} \partial_{-t'}^{\ell_2}
	\int \dd^d x'' \frac{\delta[t-t''-R_1]}{4\pi R_1^{2n+1-\ell_1}} H^{\alpha''\beta''} \frac{\delta[t''-t'-R_2]}{4\pi R_2^{2n+1-\ell_2}}, \\
R_1 &\equiv |\vec{x}-\vec{x}''|, \qquad R_2 \equiv |\vec{x}''-\vec{x}'| .
\end{align}
Integrating over $t''$ collapses the 2 $\delta$-functions into one; and averaging the two solutions for $t''$ tells us $t'' = (t+t'+R_2-R_1)/2$. At this point we switch to prolate ellipsoidal coordinates centered at $(\vec{x}+\vec{x}')/2$, i.e.,
\begin{align}
\label{ProlateEliipsoidalIntegrationCoordinates_IofII}
\vec{x}'' \equiv \frac{1}{2}(\vec{x}+\vec{x}') + \vec{x}''',
\end{align}
where
\begin{align}
\label{ProlateEliipsoidalIntegrationCoordinates_IIofII}
\vec{x}''' [s'',\theta'',\Omega''_{2n+1}]
\equiv \frac{1}{2} \sqrt{s''^2-R^2} \sin[\theta''] \widehat{n}[\Omega''_{2n+1}]
		+ \frac{s''}{2} \cos[\theta''] \widehat{z}, 
		\qquad s'' \geq R, \qquad \widehat{z} \equiv \frac{\vec{x}-\vec{x}'}{| \vec{x}-\vec{x}' |}.
\end{align}
The $z$-axis is parallel to $\vec{x}-\vec{x}'$, the spatial vector joining source to observer; $\widehat{n}$ is the unit spatial vector lying in the plane orthogonal to the $z$-axis. We have
\begin{align}
\int_{\mathbb{R}^{3+2n}} \dd^{3+2n} \vec{x}'' 
= \frac{1}{2} \int_R^\infty \dd s'' &\int_{-1}^{+1} \dd c'' \left(1-c''^2\right)^n \int_{\mathbb{S}^{2n+1}} \dd\Omega_{2n+1}^{(\widehat{n})} \nonumber\\
&\times	\left( \left(\frac{s''}{2}\right)^2 - \left(\frac{R}{2}\right)^2 \right)^n 
	\left( \left(\frac{s''}{2}\right)^2 - \left(\frac{R}{2}\right)^2 c''^2 \right) ,
\end{align}
while a direct calculation hands us
\begin{align}
R_1 = \frac{s''}{2} - \frac{R}{2} c'', \qquad
R_2 = \frac{s''}{2} + \frac{R}{2} c'' .
\end{align}
We gather, at this point,
{\allowdisplaybreaks\begin{align}
\left(G|1\right)_{x,x'} &=
\frac{1}{2(2\pi)^{2n}} \sum_{\ell_1,\ell_2=0}^{n}  A_{\ell_1}^{(n)} A_{\ell_2}^{(n)} 
	\partial_\alpha \partial_{\beta'} \partial_t^{\ell_1} \partial_{-t'}^{\ell_2} \nonumber\\
&\qquad\qquad\times
	\int_R^\infty \dd s'' \int_{-1}^{+1} \dd c'' \left(1-c''^2\right)^n \int_{\mathbb{S}^{2n+1}} \dd\Omega_{2n+1}^{(\widehat{n})}
		\left( \left(\frac{s''}{2}\right)^2 - \left(\frac{R}{2}\right)^2 \right)^n \\
&\qquad\qquad\times
\frac{\delta[t-t'-s'']}{(4\pi)^2 \left(\left(\frac{s''}{2}\right)^2 - \left(\frac{R}{2}\right)^2 c''^2\right)^{2n}} 
\left(\frac{s''}{2}-\frac{R}{2}c''\right)^{\ell_1} \left(\frac{s''}{2}+\frac{R}{2}c''\right)^{\ell_2}
H^{\alpha''\beta''}[t'',\vec{x}''] \nonumber
\end{align}}
The $\delta$-function instructs us to drop the $s''$ integration and set $s''=t-t'$. Because of the lower limit $s''=R$, however, integrating over $s''$ yields a non-zero answer only if $t-t' \geq R > 0$: the $s''$ integral is thus proportional to $\Theta[t-t'-R] = \Theta[t-t'] \Theta[\bar{\sigma}]$ with $\bar{\sigma} \equiv (1/2)((t-t')^2-R^2)$. We have arrived at the result in eq. \eqref{OrderhGreensFunction_MasslessScalar}.

\section{Static Newtonian Spacetimes}
\label{Section_Newtonian}

We will move on, in this section, to impose that the linearized Einstein's equations hold for our weak field geometry sourced primarily by a time independent, spatially localized, but otherwise arbitrary mass density $T_{00}[\vec{x}]$. This implies -- if we assume that asymptotically flat solutions to the linearized Einstein's equations are unique -- the geometry, written in pseudo-Cartesian coordinates, takes the form
\begin{align}
\label{StaticNewtonianSpacetime}
g_{\mu\nu} = \eta_{\mu\nu} + h_{\mu\nu}, \qquad
h_{00}[\vec{x}] = -2(d-3) \Phi[\vec{x}], \qquad h_{ij}[\vec{x}] = -2 \delta_{ij} \Phi[\vec{x}], \\
h_{0i} = 0, \qquad h \equiv \eta^{\mu\nu} h_{\mu\nu} = 4\Phi . \nonumber
\end{align}
This is because, with such a metric, the only non-zero component of the linearized Einstein's equation is
\begin{align}
\label{LinearizedEinstein}
G_{00} = -(d-2) \vec{\nabla}^2 \Phi[\vec{x}] = \frac{T_{00}[\vec{x}]}{4 \mpl^{d-2}}, \qquad \mpl \equiv \frac{1}{\sqrt{32 \pi \GN}} ,
\end{align}
where $\vec{\nabla}^2 = \delta^{ij} \partial_i \partial_j$ and $\GN$ is Newton's constant. Moreover, let us define the multipole moments
\begin{align}
\label{MultipoleExpansion}
M^{i_1 \dots i_q} 
&\equiv \int \dd^{d-1}\vec{x}' \frac{\Gamma \left[\frac{d-3}{2}\right]}{4 \pi ^{\frac{d-1}{2}}}
			\frac{T_{00}[\vec{x}']}{4 \mpl^{d-2}} x'^{i_1} \dots x'^{i_q} .
\end{align}
\footnote{Note that our moments are defined slightly differently from Poisson's \cite{Poisson:2002jz}. For instance, when $d=4$, our monopole is twice the mass, $M = 2 \GN \int \dd^{d-1} \vec{x}' T_{00}[\vec{x}']$.}We shall assume the mass density $T_{00}$ vanishes outside some radius $|\vec{x}| > r_c$, otherwise the spatial integrals computing the multipole moments may not converge. In terms of these moments, the tail part of the massless scalar Green's function reads
{\allowdisplaybreaks
\begin{align}
\label{G_EvenD_StaticNewtonian_Tail}
\left(G|1\right)_{x,x'}^\text{(Tail)} &=
- \frac{\Theta[t-t'] \Theta[\bar{\sigma}]}{4(2\pi)^{2n+2}} 
\sum_{q=0}^{\infty} \frac{(-)^q}{q!} M^{i_1 \dots i_q} \partial_{i_1^+} \dots \partial_{i_q^+} 
\sum_{\ell_1,\ell_2=0}^{n} A_{\ell_1}^{(n)} A_{\ell_2}^{(n)} \partial_t^{\ell_1+1} \partial_{-t'}^{\ell_2+1} 
		\left\{ \left( \frac{\bar{\sigma}}{2} \right)^n \mathcal{I}^{(\ell_1,\ell_2)} \right\} , \nonumber\\
\partial_{\mu^+} &\equiv \partial_{x^\mu} + \partial_{x'^\mu}, \qquad
\bar{\sigma} \equiv \frac{(t-t') ^2-R^2}{2}, \qquad R\equiv|\vec{x}-\vec{x}'|,
\end{align}}
with
\begin{align}
\label{MasterIntegral}
\mathcal{I}^{(\ell_1,\ell_2)}
&= \sum_{\ell=0}^{\infty} \sum_{k=1}^{\max[2n-\ell_1,2n-\ell_2]}
\frac{(-)^n \pi^{n+1} \Gamma\left[n+\frac{1}{2}\right] 2^{k+3 n-\ell+3} (k+\ell-1)!}{(k-1)! \Gamma \left[n+\ell +\frac{1}{2}\right]} \nonumber\\
&\times
\left(\mathfrak{T}_{\left(1,2 n-\ell_1\right)}[k] \binom{4n-k-\ell _1-\ell _2-1}{2n-k-\ell_1}
	+(-)^{\ell } \mathfrak{T}_{\left(1,2 n-\ell _2\right)} \binom{4n-k-\ell _1-\ell _2-1}{2 n-k-\ell _2}\right) \nonumber\\
&\times
R^{-2 n+\ell -1} (t-t')^{-4 n-\ell +\ell _1+\ell _2}
C_{\ell}^{\left(n+\frac{1}{2}\right)}\left[\frac{s _<}{R}\right] 
\left(\left(\frac{s_>}{R}\right)^2-1\right)^{-n/2} Q_{n+\ell }^{-n}\left[\frac{s _>}{R}\right]
C_{\ell}^{\left(n+\frac{1}{2}\right)}\left[\frac{r'-r}{R}\right] \nonumber\\
&\times
\,_2F_1\left[\frac{k+\ell }{2},\frac{1}{2} (k+\ell+1);n+\ell +\frac{3}{2};\left(\frac{R}{t-t'}\right)^2\right], 
\end{align}
where (the discrete ``top-hat") $\mathfrak{T}_{\left(a,b\right)}[k]=1$ for $a \leq k \leq b$ ($a$, $b$, and $k$ are integers), and $\mathfrak{T}_{\left(a,b\right)}[k]=0$ otherwise; $\binom{4n-k-\ell _1-\ell _2-1}{2n-k-\ell_{1,2}}$ are binomial coefficients; $C_\ell^{(\lambda)}$ is the Gegenbauer polynomial; $Q_\nu^\mu$ is the Legendre function of the second kind; and $\,_2F_1$ is the hypergeometric function. (We record that, via equation (8.703) of Gradshteyn and Ryzhik \cite{GS}, it is possible to re-write the $\,_2 F_1$ in terms of $Q_{n+\ell}^{k-n-1}$, radicals and powers of $R/(t-t')$.) We also define $r \equiv |\vec{x}|$ and $r' \equiv |\vec{x}'|$. At early times, $t-t' < r+r'$, we have 
\begin{align}
s_> = r+r', \qquad s_< = t-t' ;
\end{align}
whereas at late times, $t-t' > r+r'$, we have
\begin{align}
s_> = t-t', \qquad s_< = r+r' .
\end{align}
To recover from eq. \eqref{G_EvenD_StaticNewtonian_Tail} and \eqref{MasterIntegral} the $d=4$ result, simply discard the summations with respect to $\ell$ and $k$, set $\ell=0$, followed by $n=k=0$, and replace each of the two binomial coefficients with $1/2$. (We will justify this statement below.) For $t-t'>r+r'$, one would obtain
\begin{align}
\label{4D_Tail}
4D: \qquad \left(G|1\right)_{x,x'}^\text{(Tail)} 
&= -\Theta[t-t']\Theta[\bar{\sigma}] \frac{M}{\pi} \frac{t-t'}{\left((t-t')^2-|\vec{x}-\vec{x}'|^2\right)^2} .
\end{align}
The $\left(G|1\right)_{x,x'}^\text{(Tail)}$ here is in fact $\Theta[t-t']\Theta[\bar{\sigma}]/(4\pi)$ times Poisson's \cite{Poisson:2002jz} eq. (1.4), if we remember to replace $M \to 2M$ in eq. \eqref{4D_Tail}.

{\it Symmetries of $(G|1)^{\text{(Tail)}}$} \qquad The $\mathcal{I}^{(\ell_1,\ell_2)}$ in eq. \eqref{MasterIntegral} depends on the spacetime locations of the observer and the source of the Green's function, $(t,\vec{x})$ and $(t',\vec{x}')$ respectively, through the combination $t-t'$, $R$, $r$, and $r'$, i.e.,
\begin{align}
\mathcal{I}^{(\ell_1,\ell_2)} = \mathcal{I}^{(\ell_1,\ell_2)}[t-t',R,r,r'] .
\end{align}
It is invariant under global rotations and inversions of the spatial vectors, and under time translations. However, because of the explicit dependence on the radial coordinates $r$ and $r'$, it is not invariant under spatial translations, namely, under the simultaneous replacements $\vec{x} \to \vec{x} + \vec{c}$ and $\vec{x}' \to \vec{x}' + \vec{c}$ (for constant $\vec{c}$).\footnote{There is a possibility that, the series in equations \eqref{G_EvenD_StaticNewtonian_Tail} and \eqref{MasterIntegral} may be simplified or summed up, such that the explicit dependence on $r$ and $r'$ vanishes; however, we do not know how to address this question.} When acted upon by the symmetrized derivatives occurring in the higher multipole terms in eq. \eqref{G_EvenD_StaticNewtonian_Tail}, the result is non-zero. Therefore we see that, in even dimensions higher than 4, multipole moments -- which generically break spacetime rotational symmetry\footnote{For example, $M^i$ defines a particular direction in space.} -- do contribute to the late time tail of the scalar Green's function, unlike in 4D. Now, let $r_c$ denote the characteristic size of the material body sourcing the weak field geometry. As a rough estimate, if we assign the scalings $x'^i \sim r_c$ in eq. \eqref{MultipoleExpansion} and $\partial/\partial x^i \sim 1/r$, $\partial/\partial x'^i \sim 1/r'$ in eq. \eqref{G_EvenD_StaticNewtonian_Tail}, we may deduce that the contribution to the tail part of the scalar Green's function from the $q$th moment scales as
\begin{align}
\frac{1}{q!} \left(\frac{r_c}{\min[r,r']}\right)^q
\end{align}
relative to that of the monopole moment.

{\bf Derivation of eq. \eqref{G_EvenD_StaticNewtonian_Tail}} \qquad The linearized Einstein equation in eq. \eqref{LinearizedEinstein} has the solution
\begin{align}
\label{TaylorExpansion}
(d-2) \Phi 
&= \int \dd^{d-1}\vec{x}' \frac{\Gamma \left[\frac{d-3}{2}\right]}{4 \pi ^{\frac{d-1}{2}} |\vec{x}-\vec{x}'|^{d-3}}
			\frac{T_{00}[\vec{x}']}{4 \mpl^{d-2}} .
\end{align}
If we expand $|\vec{x}-\vec{x}'|^{3-d}$ about $\vec{x}' = \vec{0}$, 
\begin{align}
(d-2) \Phi &= \sum_{q=0}^{\infty} \frac{(-)^q}{q!} M^{i_1 \dots i_q} \partial_{i_1} \dots \partial_{i_q} \frac{1}{|\vec{x}|^{d-3}} .
\end{align}
Let us now observe that, in eq. \eqref{deltaG_Start}, the $H^{\alpha\beta} = (1/2) h \eta^{\alpha\beta} - h^{\alpha\beta}$ is
\begin{align*}
H^{00} = 2(d-2)\Phi, \qquad H^{ij} = H^{0i} = 0 .
\end{align*}
Inserting the expansion of eq. \eqref{TaylorExpansion} into eq. \eqref{deltaG_Start}, exploiting the property that (for an arbitrary function $F$)
\begin{align}
\label{SymmetrizedDerivatives}
\int \dd^d x'' \overline{G}_{x,x''} \partial_{\mu''} F[x''] \overline{G}_{x'',x'}
= \partial_{\mu^+} \int \dd^d x'' \overline{G}_{x,x''} F[x''] \overline{G}_{x'',x'} ,
\end{align}
followed by application of eq. \eqref{OrderhGreensFunction_MasslessScalar} then informs us that $\left(G|1\right)_{x,x'}^\text{(Tail)}$ takes the form in eq. \eqref{G_EvenD_StaticNewtonian_Tail}, and the integral that remains to be evaluated is
\begin{align}
\label{StaticNewtonian_MasterIntegral}
\mathcal{I}^{(\ell_1,\ell_2)}
\equiv 	
\int_{-1}^{+1} \dd c'' \left(1-c''^2\right)^n \int_{\mathbb{S}^{2n+1}} \dd\Omega_{2n+1}^{(\widehat{n})}
&\left(\frac{t-t'}{2}-\frac{R}{2}c''\right)^{\ell_1-2n} \left(\frac{t-t'}{2}+\frac{R}{2}c''\right)^{\ell_2-2n} \nonumber\\
&\times
 \left\vert \frac{\vec{x}+\vec{x}'}{2} + \sqrt{\frac{\bar{\sigma}}{2}} \sqrt{1-c''^2} \widehat{n} + \frac{t-t'}{2} c'' \widehat{z} \right\vert^{-2n-1} .
\end{align}
One can show that the ellipsoidal coordinates of $-(\vec{x}+\vec{x}')/2$ are
\begin{align}
\left(s_+, \cos\theta_+\right) = \left(r+r', \frac{r'-r}{R}\right) .
\end{align}
We may then, in eq. \eqref{StaticNewtonian_MasterIntegral}, invoke the prolate ellipsoidal harmonic expansion of eq. \eqref{ProlateEllipsoidalHarmonicDecomposition} -- setting $\rho \to s/2$ there. The integral over the $\mathbb{S}^{2n+1}$ eliminates every $C_m^{\left( \frac{D-3}{2} \right)}[\widehat{n}_{D-1} \cdot \widehat{n}'_{D-1}]$ except the $m=0$ term. Because $C_0^{(\lambda)}[z] = 1$, we have
\begin{align}
\label{MasterIntegral_D}
\mathcal{I}^{(\ell_1,\ell_2)}
= \Omega_{2n+1} &\int_{-1}^{+1} \dd c'' 
		\frac{\left(1-c''^2\right)^{\frac{D-3}{2}} C_\ell^{\left(\frac{D}{2}-1\right)}[c'']}{\left(\frac{t-t'}{2}-\frac{R}{2}c''\right)^{2n-\ell_1} \left(\frac{t-t'}{2}+\frac{R}{2}c''\right)^{2n-\ell_2}} \nonumber\\
&\times \frac{i^{D+1}}{(R/2)^{D-2}} \sum_{\ell=0}^\infty 
\frac{2^{\frac{1}{2} (D-3)} (D+2 \ell -2) \Gamma[D-2] \ell! \Gamma\left[\frac{D}{2}-1\right]^3}{\sqrt{\pi} \Gamma \left[\frac{D}{2}-1\right]^2 (D-3+\ell)!} \nonumber\\
&\times
	C_\ell^{\left(\frac{D}{2}-1\right)}\left[\xi_<\right] 
	\left(\xi_>^2-1\right){}^{\frac{3-D}{4}} Q_{\frac{D-3}{2}+\ell}^{-\frac{D-3}{2}}\left[\xi _>\right] 
	C_\ell^{\left(\frac{D}{2}-1\right)}\left[ \frac{r'-r}{R} \right] ,
\end{align}
for $D=3+2n$ and $\Omega_{2n+1} = 2\pi^{n+1}/n!$ is the solid angle in $2n+2$ spatial dimensions. What remains is to tackle the one dimensional integral on the first line of eq. \eqref{MasterIntegral_D}. Because $\ell_{1,2} \leq n$ and hence $2n-\ell_{1,2}$ is a positive integer, we may perform a partial fractions decomposition using the general formula
\begin{align}
\label{PartialFractions}
\frac{1}{(z-a)^\alpha (z-b)^\beta} &= 
\sum_{k=1}^\alpha
\binom{\beta+\alpha-k-1}{\alpha-k} \frac{(-)^{\alpha-k}}{
(a-b)^{\beta+\alpha-k} (z-a)^k } \nonumber\\
&\qquad\qquad
+ \sum_{k=1}^\beta
\binom{\beta+\alpha-k-1}{\beta-k} \frac{(-)^{\beta-k}}{ (b-a)^{\beta+\alpha-k}
(z-b)^k }
\end{align}
(with $\alpha,\beta=1,2,3,\dots$) followed by utilizing the integral
\begin{align}
\int_{-1}^{+1} \dd c \frac{(1-c^2)^{\lambda-(1/2)} C_\ell^{(\lambda)}[c]}{(c+\xi)^\nu}
&= 
\frac{\pi  (-)^{\ell} \Gamma[\ell +2 \lambda] \Gamma[\ell +\nu]}{2^{2 \lambda + \ell -1} \Gamma[\lambda] \Gamma[\nu] \Gamma[\ell +\lambda +1] \ell!} 
\nonumber\\
&\qquad\qquad\times
   \frac{1}{\xi^{\nu+\ell}}
   \,_2F_1\left[\frac{\ell +\nu }{2},\frac{1}{2} (\ell +\nu+1);\ell +\lambda +1;\frac{1}{\xi ^2}\right] ,
\end{align}
(where $\ell=0,1,2,3,\dots$, $\xi \geq 1$, and $\lambda + \ell>-1/2$) to achieve the result in eq. \eqref{G_EvenD_StaticNewtonian_Tail}.

Notice, when $d=4$ and thus $n=0$, the only term in eq. \eqref{MasterIntegral_D} is for $\ell_1 = \ell_2 = 0$. This in turn says we need to drop the $k$-summation and replace each of the binomial coefficients on the right hand side of the partial fractions formula in eq. \eqref{PartialFractions} with $1/2$, followed by setting $k=\alpha=\beta=n=0$.

\section{Summary and Future Directions}
\label{Section_Conclusions}

In eq. \eqref{OrderhGreensFunction_MasslessScalar} of this paper, we have worked out, up to first order in metric perturbations, the general formula for the retarded Green's function of a minimally coupled massless scalar in an arbitrary even dimensional $(d \geq 4)$ weak field geometry. We also specialized to a static Newtonian spacetime (eq. \eqref{StaticNewtonianSpacetime}) and computed the tail part of the Green's function (eq. \eqref{G_EvenD_StaticNewtonian_Tail}) in terms of the multipole moments of the mass density. Because all multipole moments appear to contribute to the Green's function tail for even $d > 4$ -- the symmetrized derivatives $\partial_{i^+}$ are not acting on an obviously space-translation symmetric object -- spacetime translation and rotational symmetries are lost even at late times, unlike the 4D case. We anticipate, therefore, that the phenomenon of tail propagation in asymptotically flat spacetimes in higher even dimensions will present richer phenomenology than the 4D situation.

In his 4D calculation, Poisson \cite{Poisson:2002jz} was able to use the Green's function he derived to evolve the scalar field forward in time. In the higher even dimensional case, this appears much more technically challenging because of the presence of the non-trivial special functions. Perhaps one may instead consider the scalar self-force problem in even dimensional weakly curved spacetimes, to see if introducing a non-trivial geometry somehow increases the sensitivity of the self-force to the detailed structure of the compact body in question. Finally, a natural extension of the work here is to compute the photon and graviton versions of the scalar Green's function results in equations \eqref{OrderhGreensFunction_MasslessScalar} and \eqref{G_EvenD_StaticNewtonian_Tail}. We hope to re-visit these issues in the future.

\section{Acknowledgments}

I thank Ofek Birnholtz, Shahar Hadar, and Eric Poisson for discussions on tails in higher dimensions, during the 17th Capra Meeting on Radiation Reaction in General Relativity at Caltech; and Shahar Hadar for questions/comments on this paper. I have employed \mma \ \cite{Mathematica} for much of the analytic calculations in this paper. This work was supported by NSF PHY-1145525 and funds from the University of Pennsylvania.

\appendix

\section{Prolate Ellipsoidal Harmonic Expansion of $|\vec{x}-\vec{x}'|^{2-D}$}
\label{Section_ProlateEllipsoidalHarmonicExpansion}

The main goal of this section is to derive the prolate ellipsoidal harmonic expansion of the Green's function $|\vec{x}-\vec{x}'|^{2-D}$ of the Laplacian in $(D \geq 3)$-dimensional flat Euclidean space. By prolate ellipsoidal coordinates, we mean that any spatial vector $\vec{x}[\rho,\theta_2,\dots,\theta_{D-1},\theta_D]$ lying in $D$ space dimensions can be expressed as
\begin{align}
\label{ProlateEliipsoidalCoordinates}
\vec{x} = \sqrt{\rho^2-R^2} \sin[\theta_D] \widehat{n}_{D-1} + \rho \cos[\theta_D] \widehat{e}_D, 
\qquad \rho \geq R > 0 .
\end{align}
Here, $\widehat{e}_D$ is the unit vector pointing parallel to the $D$th axis; $R$ is to be regarded in this section as some fixed positive constant; and the $\widehat{n}_{D-1}$ is a unit vector orthogonal to $\widehat{e}_D$, parametrized by the angles $\theta_2, \dots, \theta_{D-1}$.

The Euclidean metric is converted from $\delta_{ij} \dx{x}^i \dx{x}^j$ in Cartesian coordinates to
\begin{equation}
g_{ij}\dx{x}^i \dx{x}^j \equiv
\frac{\rho^2 - R^2 \cos^2[\theta_D]}{\rho^2 - R^2} \dx{\rho}^2
+ \left( \rho^2 - R^2 \cos^2[\theta_D] \right) (\dx{\theta}_D)^2
+ \left( \rho^2 - R^2 \right) \sin^2[\theta_D] \Omega_\text{IJ} \dx{\theta}^\text{I} \dx{\theta}^\text{J}
\end{equation}
where
\begin{align}
\Omega_\text{IJ} \dx{\theta}^\text{I} \dx{\theta}^\text{J}
    \equiv (\dx{\theta}_{D-1})^2 + \sum_{a=2}^{D-2} \sin^2[\theta_{a+1}] \dots \sin^2[\theta_{D-1}] (\dx{\theta}_a)^2
\end{align}
and
\begin{align}
|g|^{1/2} &= (\rho^2-R^2)^{\frac{D-3}{2}} \left( \rho^2 - R^2 \cos^2[\theta_D] \right) \sin^{D-2}[\theta_D] \sqrt{\Omega}, \\
\sqrt{\Omega} &\equiv \sin[\theta_3] \sin^2[\theta_4] \dots \sin^{D-3}[\theta_{D-1}] .
\end{align}
{\bf Homogeneous Solution to Laplace's Equation} \qquad We will begin by sketching the derivation for the homogeneous solution to Laplace's equation in these coordinates:
\begin{align}
\label{LaplaceEquation_D_Dimensions}
0 
= g^{ij}\nabla_i \nabla_j \psi
&= \frac{1}{\rho^2-R^2 c_D^2} \left( \frac{1}{(\rho^2-R^2)^{\frac{D-3}{2}}} \partial_\rho \left( (\rho^2-R^2)^{\frac{D-1}{2}} \partial_\rho \psi \right)
+ \frac{1}{s_D^{D-2}} \partial_D \left( s_D^{D-2} \partial_D \psi \right) \right) \nonumber\\
&\qquad\qquad
    + \frac{1}{(\rho^2-R^2) s_D^2} \frac{1}{\sqrt{\Omega}} \partial_\text{I} \left( \sqrt{\Omega} \Omega^\text{IJ} \partial_\text{J} \psi \right) ,
\end{align}
where we have defined $\{ c_i \equiv \cos[\theta_i], s_i \equiv \sin[\theta_i] \vert i = 2,3,\dots,D \}$. In fact, we shall assume that $\psi$ only depends on the three coordinates $\rho$, $\theta_{D-1}$ and $\theta_D$. Upon separating variables
\begin{align}
\psi \equiv \Psi[\xi] \Phi_D[c_D] \Phi_{D-1}[c_{D-1}], \qquad \xi \equiv \rho/R, 
\end{align}
one finds that an appropriate choice of separation constants will convert eq. \eqref{LaplaceEquation_D_Dimensions} to the following three ordinary differential equations:
{\allowdisplaybreaks\begin{align}
\frac{1}{(\xi^2-1)^{\frac{D-3}{2}}} \partial_\xi \left( (\xi^2-1)^{\frac{D-1}{2}} \partial_\xi \Psi \right)
    - \left( \ell(\ell+D-2) + \frac{m(m+D-3)}{\xi^2-1} \right) \Psi &= 0 , \\
\frac{1}{(1-c_D^2)^{\frac{D-3}{2}}} \partial_{c_D} \left( (1-c_D^2)^{\frac{D-1}{2}} \partial_{c_D} \Phi_D \right)
    + \left( \ell(\ell+D-2) - \frac{m(m+D-3)}{1-c_D^2} \right) \Phi_D &= 0 , \\
\frac{1}{(1-c_{D-1}^2)^{\frac{D-4}{2}}} \partial_{c_{D-1}} \left( (1-c_{D-1}^2)^{\frac{D-2}{2}} \partial_{c_{D-1}} \Phi_{D-1} \right)
    + m(m+D-3) \Phi_{D-1} &= 0 ,
\end{align}}
where $\ell=0,1,2,3,\dots$; and $0 \leq m \leq \ell$. The general solution that is regular over the range $-1 \leq c_{D-1}, c_D \leq +1$ then reads
\begin{align}
\label{LaplaceEquation_GeneralSolution}
\psi[\xi \equiv \rho/R, \theta_D, \theta_{D-1}]
= \sum_{\ell=0}^{\infty} \sum_{m=0}^\ell 
&\left( A_\ell^m \cdot \left( \xi^2 - 1 \right)^{m/2} C_{\ell-m}^{\left( \frac{D-2}{2} + m \right)}[\xi]
    + B_\ell^m \cdot (\xi^2-1)^{\frac{3-D}{4}} Q_{\ell + \frac{D-3}{2}}^{\frac{3-D}{2}-m}[\xi] \right) \nonumber\\
&\times \sin^m[\theta_D] C_{\ell-m}^{\left( \frac{D-2}{2} + m \right)}[\cos\theta_D] C_m^{\left( \frac{D-3}{2}\right)}[\cos\theta_{D-1}], \nonumber\\
\xi \geq 1; \ 0 \leq \theta_D \leq \pi ;& \ D \geq 3; \ \ell,m = 0,1,2,3,\dots; \ m \leq \ell .
\end{align}
For $D=3$, $0 \leq \theta_{D-1} < 2\pi$; whereas for $D>3$, $0 \leq \theta_{D-1} \leq \pi$. The $A_\ell^m$ and $B_\ell^m$ are constants, $C_\ell^{(\lambda)}$ is the Gegenbauer polynomial and $Q_\nu^\mu$ is the associated Legendre function of the second kind.

{\bf Spherical Harmonic Decomposition of $|\vec{x}-\vec{x}'|^{2-D}$} \qquad We next record the spherical harmonic expansion
\begin{align}
\label{SphericalHarmonicDecomposition}
|\vec{x}-\vec{x}'|^{2-D}
&= \frac{1}{r_>^{D-2}} \sum_{\ell = 0}^\infty \left(\frac{r_<}{r_>}\right)^\ell \frac{\Gamma[D-3]}{\Gamma\left[\frac{D-2}{2}\right]^2} \sum_{m=0}^\ell \frac{2^{2m}(\ell-m)!\Gamma\left[\frac{D-2}{2}+m\right]^2}{\Gamma[D-2+\ell+m]}(D-3+2m)
\left( \sin[\theta_D] \sin[\theta_D'] \right)^m \nonumber\\
&\qquad \qquad \times C_{\ell-m}^{\left(\frac{D-2}{2}+m\right)}[\cos\theta_D] C_{\ell-m}^{\left(\frac{D-2}{2}+m\right)}[\cos\theta_D'] C_m^{\left(\frac{D-3}{2}\right)}\left[ \widehat{n}_{D-1} \cdot \widehat{n}'_{D-1} \right],
\end{align}
where
\begin{align}
\vec{x} = r \left(\sin[\theta_D] \widehat{n}_{D-1} + \cos[\theta_D] \widehat{e}_D\right), \qquad
\vec{x}' = r' \left(\sin[\theta'_D] \widehat{n}'_{D-1} + \cos[\theta'_D] \widehat{e}_D\right),
\end{align}
and
\begin{align}
r \equiv |\vec{x}|, \qquad r' \equiv |\vec{x}'|, \qquad r_> = \max[r,r'], \qquad r_< = \min[r,r'] .
\end{align}
To derive eq. \eqref{SphericalHarmonicDecomposition}, we need to apply to
\begin{align}
|\vec{x}-\vec{x}'|^{2-D} = \frac{1}{r_>^{D-2}} \left( 1 - 2 \frac{r_<}{r_>} \cos \gamma + \left(\frac{r_<}{r_>}\right)^2 \right)^{\frac{2-D}{2}} , \\
\cos \gamma = \cos[\theta_D] \cos[\theta'_D] + \sin[\theta_D] \sin[\theta'_D] \widehat{n}_{D-1} \cdot \widehat{n}'_{D-1} ,
\end{align}
the generating function definition of the Gegenbauer polynomials in eq. (8.930) of Gradshteyn and Ryzhik \cite{GS}, followed by employing the addition formula in equation (8.934.3). Let us remark that the $D=3$ case -- see, for instance, eq. (3.70) of Jackson \cite{Jackson} -- is contained in eq. \eqref{SphericalHarmonicDecomposition}.

{\bf Prolate Ellipsoidal Harmonic Decomposition of $|\vec{x}-\vec{x}'|^{2-D}$} \qquad We are now ready to show that, for $D \geq 3$,
\begin{align}
\label{ProlateEllipsoidalHarmonicDecomposition}
&|\vec{x}-\vec{x}'|^{2-D} \\
&= 
\frac{i^{D+1}}{R^{D-2}} \sum_{\ell=0}^\infty \sum_{m=0}^{\ell}
\frac{(-)^m 2^{\frac{1}{2} (D+6 m-3)} (D+2 m-3)
   (D+2 \ell -2) \Gamma[D-3]
   (\ell -m)! \Gamma\left[\frac{D}{2}+m-1\right]^3}{\sqrt{\pi} \Gamma \left[\frac{D}{2}-1\right]^2 (D+m+\ell -3)!} \nonumber\\
&\qquad\qquad\times
	\left(\xi _<^2-1\right){}^{m/2} C_{\ell -m}^{\left(\frac{D}{2}+m-1\right)}\left[\xi_<\right] 
	\left(\xi_>^2-1\right){}^{\frac{3-D}{4}} Q_{\frac{D-3}{2}+\ell}^{-\frac{D-3}{2}-m}\left[\xi _>\right] \nonumber\\
&\qquad\qquad\times
   (\sin[\theta_D] \sin[\theta'_D])^m C_{\ell-m}^{\left(\frac{D}{2}+m-1\right)}[\cos\theta_D] C_{\ell-m}^{\left(\frac{D}{2}+m-1\right)}[\cos\theta'_D]
   C_m^{\left(\frac{D-3}{2}\right)}\left[ \widehat{n}_{D-1} \cdot \widehat{n}'_{D-1} \right]  \nonumber
\end{align}
for the prolate ellipsoidal coordinates defined in eq. \eqref{ProlateEliipsoidalCoordinates} for $\vec{x}$ and a similar one for $\vec{x}'$; and
\begin{align}
\xi_{>,<} \equiv \rho_{>,<}/R, \qquad \rho_< \equiv \min[\rho,\rho'], \qquad \rho_> \equiv \max[\rho,\rho'] .
\end{align}
Because $|\vec{x}-\vec{x}'|^{2-D}$ is the homogeneous to Laplace's equation, with respect to both $\vec{x}$ and $\vec{x}'$, when $\vec{x} \neq \vec{x}'$, we can construct it using the general solution in eq. \eqref{LaplaceEquation_GeneralSolution}. Note that the large $\xi$ limit of the solutions $\left(\xi^2-1\right)^{m/2} C_{\ell-m}^{\left(\frac{D-2}{2}+m\right)}[\xi]$ for $\ell \geq m > 0$ and the $\xi \to 1^-$ limit of the solution $\left(\xi^2-1\right)^{\frac{3-D}{4}} Q_{\ell+\frac{D-3}{2}}^{\frac{3-D}{2}-m}[\xi]$ for $D \geq 3$ and $\ell \geq m \geq 0$, are singular. On the other hand, for fixed $\vec{x}$ (or $\vec{x}'$), $|\vec{x}-\vec{x}'|^{2-D}$ is regular as $\vec{x}'$ (or $\vec{x}$) tends towards spatial infinity or when it lies on the straight line joining $\pm R \widehat{z}$ (i.e., where $\xi' \text{ or } \xi \to 1^-$). These facts inform us that $\left(\xi_>^2-1\right)^{m/2} C_{\ell-m}^{\left(\frac{D-2}{2}+m\right)}[\xi_>]$ (for $\ell > 0$) and $\left(\xi_<^2-1\right)^{\frac{3-D}{4}} Q_{\ell+\frac{D-3}{2}}^{\frac{3-D}{2}-m}[\xi_<]$ (for $\ell \geq 0$) cannot occur in its expansion. Furthermore, the $\ell=0$ terms in the expansion cannot contain the constant term $C_{\ell}^{\left(\frac{D-2}{2}\right)}[\xi] C_{\ell}^{\left(\frac{D-2}{2}\right)}[\xi'] \to 1$, because $|\vec{x}-\vec{x}'|^{2-D} \to 0$ as $\xi_> \to \infty$. Therefore, we are now lead to, for constants $\{ \chi_\ell^m \}$,
\begin{align}
\label{ProlateEllipsoidalHarmonicDecomposition_Chi}
|\vec{x}-\vec{x}'|^{2-D} 
&= 
\frac{i^{D+1}}{R^{D-2}} \sum_{\ell=0}^\infty \sum_{m=0}^{\ell}
\chi_\ell^m
	(\xi_>^2-1)^{\frac{3-D}{4}} Q_{\ell + \frac{D-3}{2}}^{\frac{3-D}{2}-m}[\xi_>] 
	\left( \xi_<^2 - 1 \right)^{m/2} C_{\ell-m}^{\left( \frac{D-2}{2} + m \right)}[\xi_<] \nonumber\\
&\times
 	       (\sin[\theta_D] \sin[\theta'_D])^m C_{\ell-m}^{\left(\frac{D}{2}+m-1\right)}[\cos\theta_D] C_{\ell-m}^{\left(\frac{D}{2}+m-1\right)}[\cos\theta'_D]
	       C_m^{\left(\frac{D-3}{2}\right)}\left[ \widehat{n}_{D-1} \cdot \widehat{n}'_{D-1} \right] .
\end{align}
At this point, we observe from eq. \eqref{ProlateEliipsoidalCoordinates} that the large $\rho_{>,<}$ limit recovers spherical coordinates from prolate ellipsoidal ones. Because the constants $\{ \chi_\ell^m \}$ in eq. \eqref{ProlateEllipsoidalHarmonicDecomposition_Chi} are independent of $\rho$, we may thus take the $\xi_{>,<} \gg 1$ limit of its $\xi$-dependent portions, and solve for $\chi_\ell^m$ by demanding that the result yields the spherical harmonic expansion in eq. \eqref{SphericalHarmonicDecomposition}.

\section{Late time tails of photons and gravitons in a 4D weakly curved asymptotically flat static spacetime}
\label{Section_LateTimeTails_4D}
In this section we build on the calculations by Poisson \cite{Poisson:2002jz} for minimally coupled massless scalars and that in \cite{Chu:2011ip}, to explain why the late time tails of photons and gravitons, in a 4D weakly curved asymptotically flat static spacetime are governed solely by the monopole moment of the source of the weakly curved geometry itself, and are thus directly related to their scalar cousin.

In equations (84), and (86) through (90) of \cite{Chu:2011ip}, the tail part of the scalar, Lorenz gauge photon and de Donder gauge graviton retarded Green's function are respectively displayed in terms of integrals involving the general metric perturbation $h_{\mu\nu}$ in eq. \eqref{WeakFieldGeometry}. If we assume, as does Poisson \cite{Poisson:2002jz}, that the weak field 4D geometry obeys the linearized Einstein's equations, then the $h_{\mu\nu}$ can be parameterized by his equations (2.1) through (2.3),
\begin{align}
h_{\mu\nu} \dd x^\mu \dd x^\nu = - 2 \Phi \delta_{\mu\nu} \dd x^\mu \dd x^\nu + 8 A_i \dd x^i \dd t,
\end{align}
with $\Phi$ and $A_i$ obeying 
\begin{align}
-\vec{\nabla}^2 \Phi = 4\pi T_{00}, \qquad -\vec{\nabla}^2 A_i = 4\pi T_{0i},
\end{align}
and the ensuing multipole expansions in his equations (2.4) and (2.5). Because every tail term is proportional to $\widehat{\mathcal{I}}_{\alpha\beta}$ (or its trace) -- which in turn, is defined in equations (78) and (79) of \cite{Chu:2011ip} -- by inserting these multipole expansions into the integral representation of $\widehat{\mathcal{I}}_{\alpha\beta}$, and exploiting eq. \eqref{SymmetrizedDerivatives}, we will deduce that each component of the tail portion of the scalar, photon and graviton Green's function can be expressed as an infinite sum of symmetrized derivatives (with respect to $x$ and $x'$) acting on the tail part of the same Green's functions computed solely from the $1/r$ moment of the 4D weak field geometry (i.e., the piece proportional to the monopole moment). For example, one such term (eq. (88) of \cite{Chu:2011ip}) reads
\begin{align}
\label{4D_RiemannTail}
\widehat{\left( R \vert 1 \right)}_{\alpha\beta \mu\nu}
= \frac{1}{2} \left( \partial_{\beta^+} \partial_{\mu^+} \widehat{\mathcal{I}}_{\nu\alpha} 
		- \partial_{\alpha^+} \partial_{\mu^+} \widehat{\mathcal{I}}_{\nu\beta} - (\mu \leftrightarrow \nu) \right),
\end{align}
where now
\begin{align}
\label{4D_I-integral}
\widehat{\mathcal{I}}_{\nu\nu} &= 
		\sum_{q=0}^\infty \frac{(-)^q}{q!} \widehat{M}^{i_1 \dots i_q} \partial_{i_1^+} \dots \partial_{i_q^+} \widehat{\mathbb{I}}_{(1)} , \qquad
		\text{(No summation over $\nu$ implied.)} \\
\widehat{\mathcal{I}}_{0j} &= \sum_{q=0}^\infty \frac{(-)^q}{q!} \widehat{J}_j^{\phantom{j}i_1 \dots i_q} \partial_{i_1^+} \dots \partial_{i_q^+}
		\widehat{\mathbb{I}}_{(1)} , 
		\qquad\qquad
\widehat{\mathcal{I}}_{ij} = 0, \qquad \text{(for $i \neq j$)}
		\nonumber
\end{align}
and
\begin{align}
\widehat{\mathbb{I}}_{(1)} &\equiv 
\int_{\mathbb{S}^2} \frac{\dd\Omega_{(\widehat{n})}}{4\pi}
\left\vert \frac{\vec{x}+\vec{x}'}{2} + \sqrt{\frac{\bar{\sigma}}{2}} \sqrt{1-c''^2} \widehat{n} + \frac{t-t'}{2} c'' \widehat{z} \right\vert^{-1} .
\end{align}
Apart from the $(4\pi) ^{-1}$, $\widehat{\mathbb{I}}_{(1)}$ is the $d=4$ version of $\mathcal{I}^{(\ell_1,\ell_2)}$ in eq. \eqref{MasterIntegral}; see also eq. \eqref{StaticNewtonian_MasterIntegral}. Up to overall multiplicative constants, which are irrelevant for the discussion at hand, the $\widehat{M}^{i_1 \dots i_q}$ and $\widehat{J}_j^{\phantom{j}i_1 \dots i_q}$ in eq. \eqref{4D_I-integral} are the 4D analogs of eq. \eqref{MultipoleExpansion}. They are also proportional to Poisson's \cite{Poisson:2002jz} $(M,M^N)$ and $(J^{ab},J^{aN})$ in his equations (4.8) and (4.9).

From eq. (106) of \cite{Chu:2011ip},
\begin{align}
\label{I_(1)}
\widehat{\mathbb{I}}_{(1)}
= |\vec{x}-\vec{x}'|^{-1} 
&\bigg(
\Theta[r+r'-(t-t')] \ln\left[ \frac{r+r'+|\vec{x}-\vec{x}'|}{r+r'-|\vec{x}-\vec{x}'|} \right]  \nonumber\\
&\qquad\qquad
		+ \Theta[t-t'-(r+r')] \ln\left[ \frac{t-t'+|\vec{x}-\vec{x}'|}{t-t'-|\vec{x}-\vec{x}'|} \right]
\bigg) .
\end{align}
Since $\widehat{\mathbb{I}}_{(1)}$ is spacetime translation symmetric for $t-t' > r+r'$, and therefore acting $\partial_{i^+}$ on it gives zero, we see that the massless scalar, photon and graviton Green's function tail receives no correction from the higher multipole moments for late times. 

Moreover, in \S V of \cite{Chu:2011ip}, it has been already noted that at first order in mass $M$ and spin $\vec{S}$ in a weak field Kerr geometry, it is the spacetime translation symmetric scalar tail $\widehat{\mathcal{I}}^{(S)}$ (see eq. \eqref{4D_Tail}; and, in \cite{Chu:2011ip}, equations (116), (129), and Fig. 3), that governs the tail behavior at late times, for the minimally coupled massless scalar, Lorenz gauge photon and de Donder gauge graviton Green's function. Since we have just uncovered the fact that the late time tails of these same photons and gravitons in a weakly curved asymptotically flat static spacetime are determined, when $t-t'>r+r'$, solely by the monopole mass, that means the results in \S V of \cite{Chu:2011ip} carry directly over to the case at hand.

\end{document}